\documentstyle[preprint,aps,tighten]{revtex}

\begin{document}
\draft
\title{Screened Coulomb potentials for astrophysical nuclear fusion reactions}
\author{Theodore E. Liolios $^{1,2,3}$ \footnote{theoliol@physics.auth.gr}}

\address{$^1$European Center for Theoretical Studies in Nuclear Physics and Related Areas\\
Villa Tambosi, I-38050 Villazzano (Trento), Italy\\
\footnote{Correspondence address}$^2$University of Thessaloniki, Department of Theoretical Physics\\
Thessaloniki 54006, Greece\\
$^3$ Hellenic War College, BST 903, Greece\\
}

\maketitle

\begin{abstract}
The electron-screening acceleration of laboratory fusion reactions at
astrophysical energies is an unsolved problem of great importance to
astrophysics. That effect is modeled here by considering the fusion of
hydrogen-like atoms whose electron probability density is used in Poisson's
equation in order to derive the corresponding screened Coulomb potential
energy. That way atomic excitations and deformations of the fusing atoms can
be taken into account. Those potentials are then treated semiclassically in
order to obtain the screening (accelerating) factor of the reaction. By
means of the proposed model the effect of a superstrong magnetic field on
laboratory Hydrogen fusion reactions is investigated here for the first time
showing that, despite the considerable increase in the cross section of the $%
dd$ reaction, the $pp$ reaction is still too slow to justify
experimentation. The proposed model is finally applied on the $H^{2}\left(
d,p\right) H^{3}$ fusion reaction describing satisfactorily the experimental
data although some ambiguity remains regarding the molecular nature of the
deuteron target. Notably, the present method gives a sufficiently high
screening energy for Hydrogen fusion reactions so that the take-away energy
of the spectator nucleus can also be taken into account.
\end{abstract}


\oddsidemargin -0.25cm \evensidemargin -0.25cm \topmargin -1.0cm \textwidth %
16.3cm \textheight 22.3cm

\section{Introduction}

At astrophysical energies of a few $keV$ corresponding to stellar
temperatures of several millions degrees kelvin the cross section $\sigma
\left( E\right) $ of the predominant $s$-wave fusion reactions is given by 
\begin{equation}
\sigma \left( E\right) =\frac{S\left( E\right) }{E}P\left( E\right)
\end{equation}
where the astrophysical factor $S\left( E\right) $ embodies all the nuclear
effects of the reaction and for non-resonant cases is a slowly varying
function of the center-of-mass energy $E.$ On the other hand , the
penetrability factor $P\left( E\right) $ embodies all atomic effects of the
reaction and when the electron cloud around the fusing nuclei is ignored it
is given by $P\left( E\right) =\exp \left( -2\pi n\right) $ where $n$ is the
Sommerfeld parameter.

As the astrophysical factor varies slowly with energy we usually replace it
with a truncated Taylor series which will be studied extensively in the
present paper 
\begin{equation}
S\left( E\right) =S\left( 0\right) +S^{^{\prime }}\left( 0\right)
E+0.5S^{^{\prime \prime }}\left( 0\right) E^{2}  \label{polynomial}
\end{equation}

Any error in the zero-energy astrophysical factor $S\left( 0\right) $ is
actually an error in the corresponding reaction rate in the stellar plasma,
which in turn reflects linearly on the energy production rate.

In the past years there have been exhaustive efforts to extend measurements
of the $S\left( E\right) $ towards even lower energies\cite{kraussdd}\cite
{krausshe3} in order to obtain a reliable value for $S\left( 0\right) $.
This is necessary as extrapolating higher energy data to zero energies
introduces an inevitable numerical error. However, at such low energies, the
electron cloud that screens the fusing nuclei enhances the fusion reaction
by lowering the Coulomb barrier. Consequently, disregarding its presence
leads to an overestimation of $S\left( 0\right) $. Unfortunately, even very
recent experiments\cite{gsasso} cannot explain the screening enhancement
which exceeds all the available theoretical predictions as was recently
admitted \cite{rolfs2000}, \cite{liolios2000} .

Various authors have studied the influence of the atomic cloud on the cross
section of low energy nuclear reaction. A qualitative study\cite{assen},
which parametrized various atomic processes such as molecular formation,
excitation and ionization, yielded a fair approximation for the possible
contributions of the electronic degrees of freedom in the nuclear collision
experiment. Moreover, by assuming a constant charge density around the
target nucleus,  a subsequent model\cite{bencze} predicted a screening shift
which was compatible with the experimental data $.$ However that assumption
is an oversimplification which will be amended in the present paper. The
most sophisticated approach has been a few-body treatment\cite{bracci} which
established a lower (sudden) and a higher (adiabatic) limit for the
screening energy transferred into the relative nuclear motion. Although more
studies followed\cite{braccidd}\cite{shoppaatomic}, which also extended the
calculations to molecular fusion reactions\cite{shoppamolecular}, despite
their mathematical rigor they could not explain the discrepancy between
experimental and theoretical screening energies.

In this work there is presented a mean-field model for the study of screened
nuclear reactions at astrophysical energies in the laboratory \cite
{lioliosgreece}. That model agrees well with the available experimental
data, thus enabling us to improve the accuracy of the associated
astrophysical factor. Moreover, by means of the proposed model the effect of
a superstrong magnetic field on laboratory Hydrogen fusion reactions is also
investigated for the first time, yielding the associated magnetic
accelerating factor. Notably, the present method gives a sufficiently high
screening energy for Hydrogen fusion reactions so that the spectator nucleus
take-away energy can also be taken into account.

\section{Screened Coulomb potentials}

After the pioneering work\cite{assen} that established the importance of
atomic effects in low energy nuclear reactions various authors have tried to
create models that account for the observed enhancement. A simple model\cite
{bencze}, suggested at an early stage, assumed that the electronic charge
density around the target nucleus is constant, thus predicting for the
nucleus-atom reaction between the atomic target $Z_{1}e\,$and the projectile 
$Z_{2}e$ a screening energy $U_{e}=\left( 3/2\right) Z_{1}Z_{2}e^{2}a^{-1}.$
In order to take into account the dependence of the screening radius on the
charge state of the reaction participants, that model used a screening
radius taken from scattering experiments\cite{lind} so that 
\begin{equation}
a=0.8853a_{0}\left( Z_{1}^{2/3}+Z_{2}^{2/3}\right) ^{-1/2}
\label{screeningradius}
\end{equation}
where $a_{0}$ the Bohr radius. Although that screening energy is larger than
the one predicted by the simple formula\cite{assen} $U_{e}=Z_{1}Z_{2}e^{2}%
\left( a_{0}/Z_{1}\right) ^{-1}$ it has some very obvious defects. The
assumption that the charge density is constant leads to an unnaturally sharp
cut-off at a distance $r=a\,$from the center of the target nuclei, which is
not born out either by theory or experiment. Moreover, atomic excitations
and deformations of the target atom are totally disregarded. On the other
hand normalizing the charge distribution so that the total charge is $%
-Z_{1}e $ gives a charge density 
\begin{equation}
\rho _{0}=-\frac{3}{4}\frac{Z_{1}e}{\pi a^{3}}
\end{equation}
In order to assess the validity of that density we can consider the
hydrogen-like atom $Z_{1}e$ which will also be used in this section . The
charge density at the center of the cloud of such an atom (when the electron
is in its ground state) is $\rho _{0}^{H}=$ $-e\left( Z_{1}/a_{0}\right)
^{3}/\pi .$ It is obvious that for $Z_{1}=Z_{2}=1$ we obtain $\rho
_{0}\simeq -\left( e/a_{0}^{3}\right) $ and $\rho _{0}^{H}=-\left(
e/a_{0}^{3}\right) /\pi ,$ that is the simplified model in question
overestimates the central density by a factor of $\pi .$

Consequently it is obvious that if low energy nuclear reactions are to be
treated by means of a mean-field potential a more sophisticated treatment is
necessary.

As a first step we consider a more plausible charge distribution: 
\begin{equation}
\rho \left( r\right) =\rho _{0}\left( 1-\frac{r^{2}}{a^{2}}\right)
\label{ro1}
\end{equation}
which takes into account the depletion of charge with respect to distance
from the center. The radius $a$ is the screening radius given by Eq. $\left( 
\ref{screeningradius}\right) $ and the charge density $\rho _{0}$ at the
center of the cloud can be found by means of the normalization condition $:$ 
\begin{equation}
\int_{0}^{a}\rho \left( r\right) 4\pi r^{4}dr=-Z_{1}e  \label{norm}
\end{equation}
This integral yields a central value of 
\begin{equation}
\rho _{0}=-\frac{15}{8}\frac{Z_{1}e}{\pi a^{3}}
\end{equation}
Note that for a collision $Z_{1}=Z_{2}=1$ we have a central charge density $%
\rho _{0}=7.68\left( e/a_{0}^{3}\right) /\pi $ which gives an even larger
core density than the constant density assumption. An alternative approach
would be to consider the value $\rho _{0}$ equal to the corresponding
hydrogen-like one and then calculate the screening radius using Eq. $\left( 
\ref{norm}\right) .$ The latter treatment gives a screening radius 
\begin{equation}
a=\left( \frac{15}{8Z_{1}^{2}\pi }\right) ^{1/3}a_{0}  \label{radius2}
\end{equation}
which is independent of the charge of the projectile. For hydrogen isotopes
Eq. $\left( \ref{radius2}\right) $ gives a radius of $a=0.842a_{0}$

We can calculate the electrostatic energy by solving the equation of Poisson
for the above charge distribution with the appropriate boundary conditions,
so that 
\begin{equation}
\Phi \left( r\right) =-\frac{15}{12}\frac{Z_{1}e}{a}\left[ \frac{3}{2}%
-\left( \frac{r}{a}\right) ^{2}+\frac{3}{10}\left( \frac{r}{a}\right)
^{4}\right]  \label{fsimple}
\end{equation}
Whenever a bare nucleus $Z_{2}e$ impinges on the target nuclei surrounded by
the electron cloud of Eq.$\left( \ref{ro1}\right) \,$the total interaction
potential in the atom-nucleus reaction channel is 
\begin{equation}
V\left( r\right) =\frac{Z_{1}Z_{2}e^{2}}{r}-\frac{15}{12}\frac{%
Z_{1}Z_{2}e^{2}}{a}\left[ \frac{3}{2}-\left( \frac{r}{a}\right) ^{2}+\frac{3%
}{10}\left( \frac{r}{a}\right) ^{4}\right]  \label{v2}
\end{equation}
Although the above potential energy is more plausible than the constant
charge density one, a more reliable charge distribution should be considered
which could account for various other atomic effects as well as for the
atom-atom reaction channel.

Let us consider a hydrogen-like atom with atomic number $Z_{1}$. When the
wave function of the electron is given by $\Psi _{nl}\left( r,\theta \right) 
$ then the charge density around the point-like nucleus is 
\begin{equation}
\rho \left( r,\theta \right) =-e\left| \Psi _{nl}\left( r,\theta \right)
\right| ^{2}  \label{dist}
\end{equation}
by which it is obvious that both the previous screening model and that of
Ref. \cite{bencze} are imperfect. If we solve the equation of Poisson for
hydrogen atoms (or hydrogen-like ions) whose electron is in its ground $%
\left( 1s\right) $ state we obtain 
\begin{equation}
\Phi _{00}\left( r\right) =-\frac{e}{r}+\frac{e}{r}\left( 1+\frac{r}{2r_{0}}%
\right) \exp \left( -r/r_{0}\right) 
\end{equation}
where the screening radius is 
\begin{equation}
r_{0}=\frac{a_{0}}{2Z_{1}}
\end{equation}
If a positive projectile $Z_{2}e$ interacts with the above screened nucleus
then the total potential energy is 
\begin{equation}
V_{00}\left( r\right) =\frac{Z_{1}Z_{2}e^{2}}{r}-\frac{Z_{2}e^{2}}{r}+\frac{%
Z_{2}e^{2}}{r}\left( 1+\frac{r}{2r_{0}}\right) \exp \left( -\frac{r}{r_{0}}%
\right)   \label{v00}
\end{equation}

On the other hand if we assume that the electron is in an excited state $%
\left( 2s\right) $ then the potential energy is found to be: 
\begin{equation}
V_{10}\left( r\right) =\frac{Z_{1}Z_{2}e^{2}}{r}-\frac{Z_{2}e^{2}}{r}+\frac{%
Z_{2}e^{2}}{r}\left( 1+\frac{3}{8}\frac{r}{r_{0}}+\frac{r^{2}}{16r_{0}^{2}}+%
\frac{r^{3}}{64r_{0}^{3}}\right) \exp \left( -\frac{r}{2r_{0}}\right) 
\label{f10}
\end{equation}

It should be emphasized that in the derivation of the above potentials we
have assumed an unperturbed wavefunction of the target nuclei, throughout
the tunnelling process. In fact at astrophysical energies the electron cloud
responds rapidly and by the time tunneling begins the nuclei are so close
that the wavefunction is actually that of a hydrogen-like atom with charge $%
Z_{1}^{*}=\left( Z_{1}+Z_{2}\right) $ and a screening radius $%
r_{0}^{*}=a_{0}/2Z_{1}^{*}$

\section{Nuclear reactions at astrophysical energies}

At astrophysical energies reactions between light nuclei take place via $s$%
-interactions, thus enabling us to investigate them by means of the WKB.

If we assume that a bare nucleus $Z_{2}e$ collides at very low energy $E$
with a screened nucleus whose electron is in its ground state then the
tunneling probability according to the WKB method is: 
\begin{equation}
P\left( E\right) =\exp \left[ -\frac{2\sqrt{2\mu }}{\hbar }%
\int_{R}^{r_{c}\left( E\right) }\sqrt{V_{00}\left( r\right) -E}dr\right]
\label{pe}
\end{equation}
We can assume that the lower limit of the WKB integral is given in terms of
the mass number $A$ of the reacting nuclei : $R=1.4\left(
A_{1}^{1/3}+A_{2}^{1/3}\right) .$ For most practical purposes this lower
bound is set equal to zero as all the nuclear effects of the fusion reaction
are included in the cross section factor.

The classical turning point can be obtained by equating the relative
collision energy $E$ with the potential energy of the interaction. The
collision energy is set equal to the Gamow peak of the corresponding
reaction in the plasma so that: 
\begin{equation}
V_{00}\left( r_{c}\right) =1.220\cdot \left(
Z_{1}^{2}Z_{2}^{2}AT_{6}^{2}\right) ^{1/3}\,\,keV  \label{rc}
\end{equation}
where $A$ the reduced mass number and $T_{6}$ the temperature in million
degrees kelvin. For a wide range of light nuclei we have performed extensive
numerical solutions for Eq. $\left( \ref{rc}\right) $ as well as numerical
integrations of Eq. $\left( \ref{pe}\right) .$ At astrophysical energies,
just as is the case with the Debye-H\"{u}ckel model in plasma conditions\cite
{lioliosprc}, the results indicate that throughout the potential barrier the
potential energy $V_{00}\left( r\right) $ of Eq. $\left( \ref{v00}\right) $
can be safely replaced by the much simpler formula: 
\begin{equation}
V_{00}\left( r\right) \simeq \frac{Z_{1}Z_{2}e^{2}}{r}-\frac{%
Z_{1}^{*}Z_{2}e^{2}}{a_{0}}
\end{equation}
Therefore the WKB penetration factor can be written as: 
\begin{equation}
P\left( E\right) =\exp \left[ -\frac{2\sqrt{2\mu }}{\hbar }%
\int_{R}^{r_{c}\left( E\right) }\sqrt{\frac{Z_{1}Z_{2}e^{2}}{r}-\frac{%
Z_{1}^{*}Z_{2}e^{2}}{a_{0}}-E}dr\right]
\end{equation}
The equation for the classical turning point is modified accordingly: 
\begin{equation}
\frac{Z_{1}Z_{2}e^{2}}{r_{c}}=1.220\cdot \left(
Z_{1}^{2}Z_{2}^{2}AT_{6}^{2}\right) ^{1/3}\,\,keV
\end{equation}
where we have ignored the screening shift given by: 
\begin{equation}
U_{e}=\frac{Z_{1}^{*}Z_{2}e^{2}}{a_{0}}  \label{myshift}
\end{equation}
It is now obvious that the relative energy of the reaction has been
increased by $U_{e}$. In that case the penetration factor can be easily
found to be\cite{shoppaatomic}: 
\begin{equation}
f_{1s}\left( E\right) \simeq \exp \left[ \pi n\left( E\right) \frac{U_{e}}{E}%
\right]  \label{fue}
\end{equation}
where the subscripts indicate the excitation state of the target atom. If we
follow the same methodology for the $2s$ state we obtain 
\begin{equation}
f_{2s}\left( E\right) \simeq \exp \left[ \pi n\left( E\right) \frac{U_{e}}{4E%
}\right]  \label{fue2s}
\end{equation}
The much simpler potential model of Eq. $\left( \ref{v2}\right) $ gives a
screening factor: 
\begin{equation}
f_{0}\left( E\right) \simeq \exp \left[ \pi n\left( E\right) \frac{%
\widetilde{U_{e}}}{E}\right]
\end{equation}
with an energy shift of 
\begin{equation}
\widetilde{U_{e}}=\frac{15}{8}\frac{Z_{1}Z_{2}e^{2}}{a}  \label{uesimple}
\end{equation}
where $a$ is given either from Eq. $\left( \ref{screeningradius}\right) $ or
Eq. $\left( \ref{radius2}\right) $

\section{Magnetically catalyzed screening}

By now it is obvious that any shift $U_{e}<<E$ of the interaction potential
energy $V\left( r\right) $%
\begin{equation}
V\left( r\right) =\frac{Z_{1}Z_{2}e^{2}}{r}-U_{e}
\end{equation}
accelerates the fusion cross section of hydrogen isotopes by a factor $%
f_{1s}\left( E\right) $ given by Eq. $\left( \ref{fue}\right) .$ That
observation will prove very useful in the study of the effects of a
superstrong magnetic field on laboratory hydrogen fusion reactions which
follows.

As a matter of fact under such extreme conditions the electron-screening
cloud is deformed in the sense that it becomes compressed perpendicular and
parallel to the magnetic field so that the screening potential energy for
the strongly magnetized hydrogen atom is\cite{heyl} 
\begin{equation}
U_{e}\left( \rho ,z;\alpha \right) =\frac{e^{2}}{\widehat{\rho }}\frac{1}{%
\sqrt{2\pi }}\int_{0}^{\infty }\frac{\exp \left[ -\frac{1}{2}\left( \frac{%
\overline{\rho }^{2}}{1+u}+\frac{\overline{z}^{2}}{\alpha ^{2}+u}\right)
\right] }{\left( 1+u\right) \sqrt{\alpha ^{2}+u}}du  \label{vsc}
\end{equation}
where $\rho ,z$ are the coordinates in a cylindrical frame of reference
whose origin coincides with the point-like nucleus of the hydrogen atom.

The natural length unit in the above formula is of course the cyclotron
radius so that $\overline{\rho }=\rho /\widehat{\rho },\,\overline{z}=z/%
\widehat{\rho }$ , and $\alpha $ is a parameter which depends on the
magnetic field and is determined by the variational method. The above
formula was shown to be reliable for very strong fields whereas it becomes
inaccurate below the threshold of the intense magnetic field regime given
by: 
\begin{equation}
B_{IMF}=4.7\times 10^{9}G.
\end{equation}
In Ref. \cite{heyl} potential $\left( \ref{vsc}\right) $ was applied at zero
relative energies in order to obtain the mean-life times of hydrogen
isotopes in neutron star surfaces. However, a more recent work\cite
{lioliosdeform} used that potential in a problem where the relative energies
were of the order of $keV$ showing that for energies $E>0.5keV$ and fields
of the order of $B_{12}=0.047$ ($B_{12}$ being the field measured in $%
10^{12}G)$ the classical turning point is so deep inside the cloud that the
screening shift can be considered constant and equal to the value of the
potential at the center of the cloud given in Ref. \cite{heyl} 
\begin{equation}
U_{e}\left( 0,0;\alpha \right) =\frac{e^{2}}{\widehat{\rho }}\frac{2}{\sqrt{%
2\pi }}\frac{\ln \left( \alpha +\sqrt{\alpha ^{2}-1}\right) }{\sqrt{\alpha
^{2}-1}}  \label{fa}
\end{equation}
In the present work that approximation has been tested for various other
fields and energies. The results show that for fields as high as $B_{12}=4.7$
and interaction energies $E>0.5\,keV$ the screening effect is independent of
the angle at which the projectile enters the electron cloud and can be
considered equal to Eq. $\left( \ref{fa}\right) .$

Therefore if the target hydrogen nuclei are in such a magnetic field the
reaction is going to be accelerated by a factor 
\begin{equation}
f_{1s}\left( E\right) \simeq \exp \left[ \pi n\left( E\right) \frac{%
U_{e}\left( 0,0;\alpha \right) }{E}\right]  \label{fueb}
\end{equation}
Figures 1 and 2 depict the acceleration of the $pp$ and $dd$ reactions
respectively for various magnetic fields and interaction energies.
Especially for the $pp$ reaction it is obvious that even in such a strong
field the cross section is still significantly small. Namely, as the
corresponding zero energy astrophysical factor is $S_{pp}\left( 0\right)
\simeq 4\times 10^{-22}keV-barns,$ the screening effect in a superstrong
field $B_{12}=4.7$ can only increase $S_{pp}\left( 0\right) $ by roughly one
order of magnitude compared to the unmagnetized case.

The $dd$ reaction, on the other hand, can be significantly affected by such
a magnetic field as it is already much faster than the $pp$ one. At very low
energies the increase can be as high as two orders of magnitude compared to
the unmagnetized case.

\section{The astrophysical factor of $d-D$ nuclear reactions.}

Despite the fact that the reactions $H^{2}\left( d,p\right)
H^{3},H^{2}\left( d,n\right) He^{3}$ have been investigated since the early
days of accelerators\cite{mcneill}\cite{davidenko}\cite{arnold}, the effect
of screening on the associated astrophysical $S\left( E\right) ,$ which will
eventually be used in theoretical calculations, is still under
investigation. In the discussion that follows we will show that our model is
compatible with the experimental data of that reaction.

The appropriate treatment of a low-energy experiment should take into
account screening effects in order to calculate the respective values of $%
S\left( E\right) .$ As a matter of fact once a screening model and the
associated screening energy $U_{e}$ are adopted the corrected bare-nucleus
astrophysical factor of the experiment is actually given by 
\begin{equation}
S_{b}\left( E\right) =E\sigma \left( E\right) \exp \left( 2\pi n\right) \exp
\left( -\pi n\frac{U_{e}}{E}\right)  \label{sbcorrect}
\end{equation}
Then Eq. $\left( \ref{polynomial}\right) $ is fitted to the data corrected
through Eq. $\left( \ref{sbcorrect}\right) $ in order to obtain the
zero-energy coefficient $S\left( 0\right) .$

Any effort to extrapolate from higher-energy data or fit all the uncorrected
data with formula $\left( \ref{polynomial}\right) \,$is bound to induce
errors.

There are three different ways to analyze low energy fusion data \cite
{gsasso} which must of course be consistent with each other. We will apply
those methods on the available data\cite{greife} for $dd$ reactions ($%
E>2keV) $ and compare them with the analytic model proposed in the present
paper. First we note that for energies $E>20keV$ any screening correction is
meaningless since the exponential term of Eq. $\left( \ref{sbcorrect}\right) 
$ is very close to unity at such high energies. Therefore we can obtain the
asymptotic behavior of the astrophysical factor by using the available
high-precision experimental data\cite{brown} for higher energies which
yielded 
\begin{equation}
S_{b}\left( E\right) =55.49\left( 0.46\right) +0.094\left( 0.0054\right) E
\end{equation}
We can now reasonably assume that this should be a fair approximation of the
bare-nucleus astrophysical provided its use consistently describes the
low-energy experimental data. In fact the screened value of $S\left(
E\right) $ will now be given by 
\begin{equation}
S\left( E\right) =\left( 55.49+0.094E\right) \exp \left( \pi n\frac{%
U_{e}^{as}}{E}\right)  \label{sas}
\end{equation}
where the screening energy $U_{e}^{as}$ is determined by fitting Eq. $\left( 
\ref{sas}\right) $ to the uncorrected data of Ref. \cite{greife}, so that $%
U_{e}^{as}=0.019\left( 0.003\right) \,keV$ with $\chi ^{2}=0.028.$

The second method which will corroborate the validity of the proposed models
entails fitting all four parameters $S\left( 0\right) ,S^{^{\prime }}\left(
0\right) ,S^{^{\prime \prime }}\left( 0\right) ,U_{e}$ simultaneously to the
uncorrected experimental data. Thus we obtain a screening energy of $%
U_{e}^{all}=0.017\left( 0.003\right) $\thinspace $keV$and a bare nucleus
astrophysical factor: 
\begin{equation}
S_{b}\left( E\right) =54.54\left( 1.39\right) +0.608\left( 0.265\right)
E-0.026\left( 0.026\right)  \label{sall}
\end{equation}
with $\chi ^{2}=0.011.$ Obviously, the two previous approaches give results
which are compatible with each other as expected. Figure 3 shows that both
the previous two fits provide a satisfactory description of the screening
effect.

The third method is a straightforward application of the theoretical models
derived in the present paper. However, in order to apply those models on the
experimental data we have to take into account that the data refer to a
molecular target while our models refer to atomic ones. Hence, we have to
allow for the energy which will be carried away by the spectator nuclei plus
the reduction due to the molecular binding energy. Although this assumption
has been argued against\cite{shoppamolecular}, the actual energy reduction
for a deuteron molecular target has been calculated\cite{engstler} by a
Coulomb explosion process to be of the order of $44\,eV.$ Therefore
modifying our models for a molecular deuteron target we derive a screening
energy $U_{e}=0.010keV$ (Eq. $\left( \ref{myshift}\right) )$ and $\widetilde{%
U_{e}}=0.016\,keV$ (Eq. $\left( \ref{uesimple}\right) )$ which are in
reasonably good agreement with the experimentally obtained values. We can
now fit the formula 
\begin{equation}
S\left( E\right) =\left[ S\left( 0\right) +S^{^{\prime }}\left( 0\right)
E+0.5S^{^{\prime \prime }}\left( 0\right) E^{2}\right] \exp \left( \pi n%
\frac{U_{e}}{E}\right)
\end{equation}
by using the screening shift of our models. The results are as follows

$U_{e}=0.010$

\begin{equation}
S_{b}\left( E\right) =57.3\left( 0.41\right) +0.160\left( 0.125\right)
E-0.0056\left( 0.002\right) E^{2}
\end{equation}
with $\chi ^{2}=0.013$ and

$\widetilde{U_{e}}=0.016$

\begin{equation}
S_{b}\left( E\right) =54.93\left( 0.38\right) +0.537\left( 0.1149\right)
E-0.0225\left( 0.007\right) E^{2}
\end{equation}
with $\chi ^{2}=0.011$

Although our models are fairly compatible with the experiment there is an
inevitably degree of uncertainty in the associated astrophysical factors due
to the actual amount of energy that is carried away by the spectator nuclei
of the molecular target. In any case the models proposed here turn out to
provide a simple and effective way of describing fusion reactions between
hydrogen-like atoms.

\section{Conclusions}

This work proposes a simple and efficient model for the study of the
screening enhancing effect on low-energy nuclear fusion reactions. In that
model, the fusing atoms are considered hydrogen-like atoms whose electron
probability density is used in Poisson's equation in order to derive the
corresponding screened Coulomb potential energy. This way atomic excitations
and deformations of the reaction participants can be taken into account. The
derived mean-field potentials are then treated semiclassically, by means of
the WKB, in order to derive the screening enhancement factor which is shown
to be compatible with the experimentally obtained one for the $H^{2}\left(
d,p\right) H^{3}$ reaction, although some ambiguity remains regarding the
molecular nature of the deuteron target. Moreover, by means of the proposed
model the effect of a superstrong magnetic field on laboratory Hydrogen
fusion reactions is investigated for the first time showing that despite the
remarkable increase in the cross section of the $dd$ reaction, the $pp$
reaction is still too slow to justify experimentation.

{\bf ACKNOWLEDGMENTS}

This work was financially supported by the Greek State Grants Foundation
(IKY)  under contract \#135/2000. The revised version was written at ECT$^{*}$ during a nuclear physics
fellowship. The author would like to thank the director of ECT$^{*}$
Prof.Malfliet for his kind hospitality and support.

Prof. C.Rolfs' advice during the Bologna-2000 conference has proved
invaluable to the completion of this paper.

{\bf FIGURE CAPTIONS}

Figure 1. The screening (acceleration) factor $f_{1s}$ with respect to the
relative interaction energy of two fusing protons for various superstrong
magnetic fields (in units of $10^{12}G)$

Figure 2. The screening (acceleration) factor $f_{1s}$ with respect to the
relative interaction energy of two fusing deuterons for various superstrong
magnetic fields (in units of $10^{12}G)$

Figure 3. The $H^{2}\left( d,p\right) H^{3}$ astrophysical factor $S\left(
E\right) $ measured in keV-b with respect to the center of mass interaction
energy $E_{cm}$ $\left( keV\right) .$ The data (squares) are taken from Ref. 
\cite{greife}$.\,$\thinspace The solid curve represents Eq. $\left( \ref{sas}%
\right) $, which makes use of the asymptotic form given in Ref. \cite{brown}%
. The dashed curve represents Eq. $\left( \ref{sall}\right) $ where all four
parameters $S\left( 0\right) ,S^{^{\prime }}\left( 0\right) ,S^{^{\prime
\prime }}\left( 0\right) ,U_{e}$ are fitted simultaneously. The dotted curve
is obtained by adopting as a screening energy the value given by Eq. $\left( 
\ref{uesimple}\right) ,$ while the dash-dotted curve stands for the
astrophysical factor obtained by using Eq. $\left( \ref{myshift}\right) .$

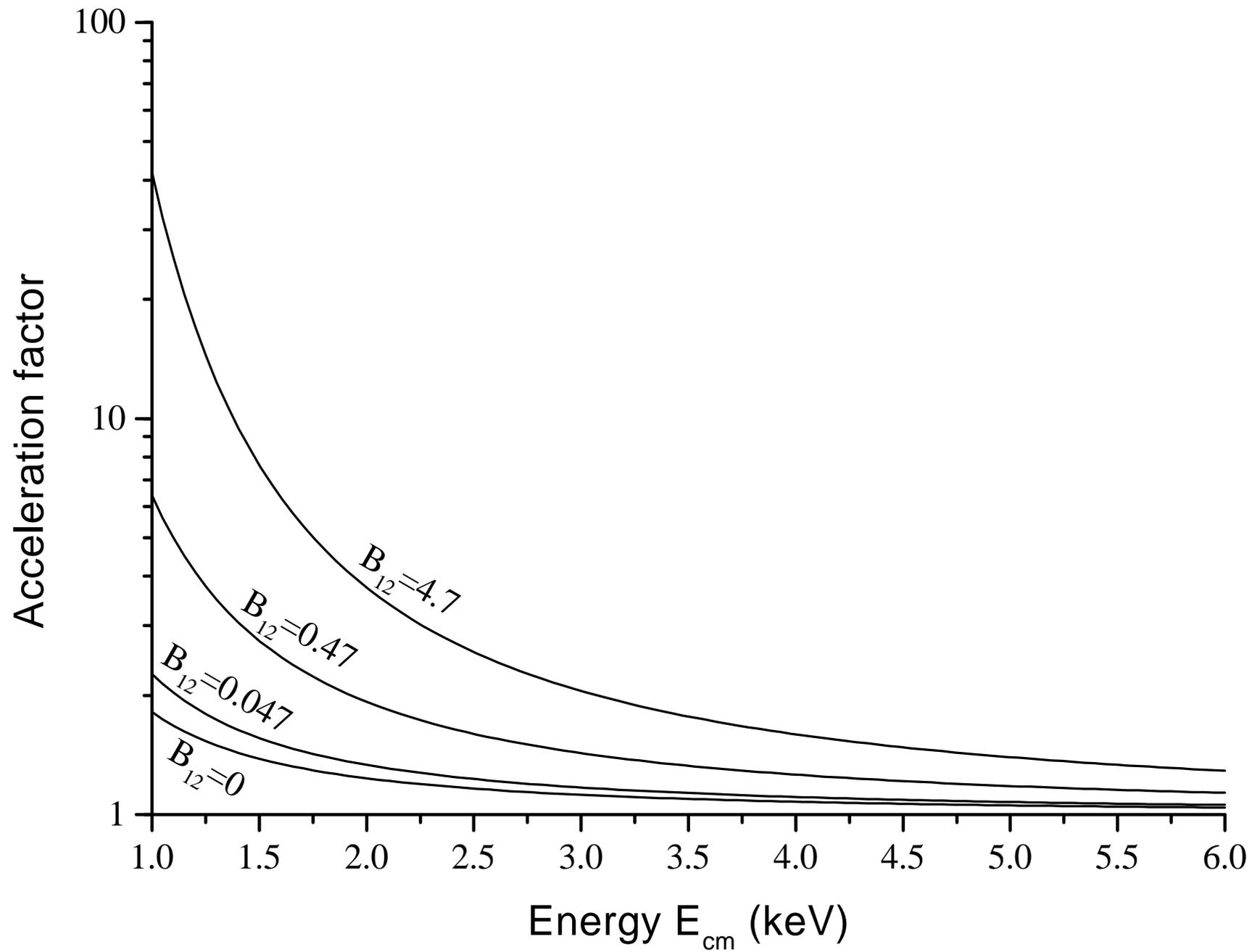
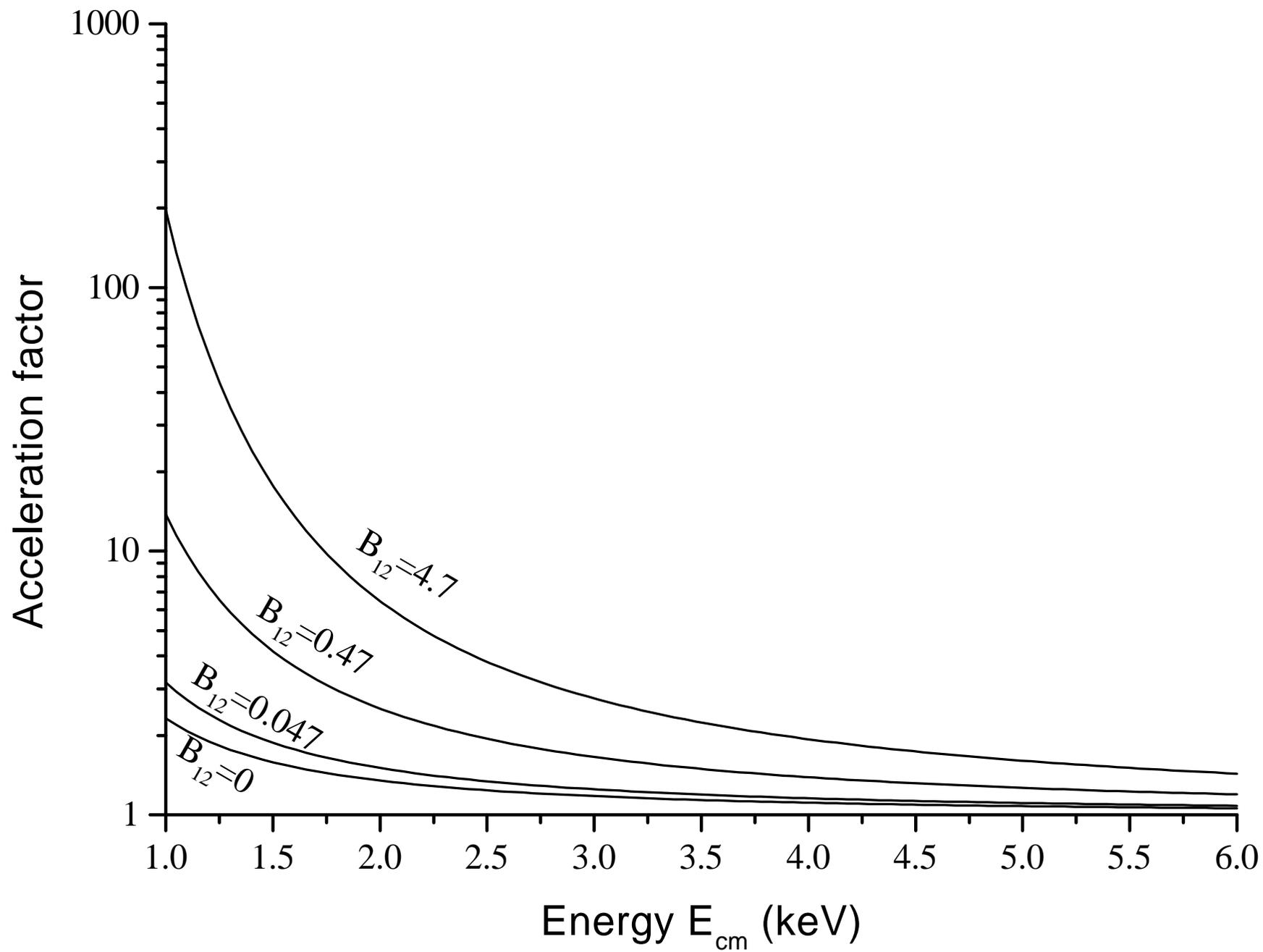
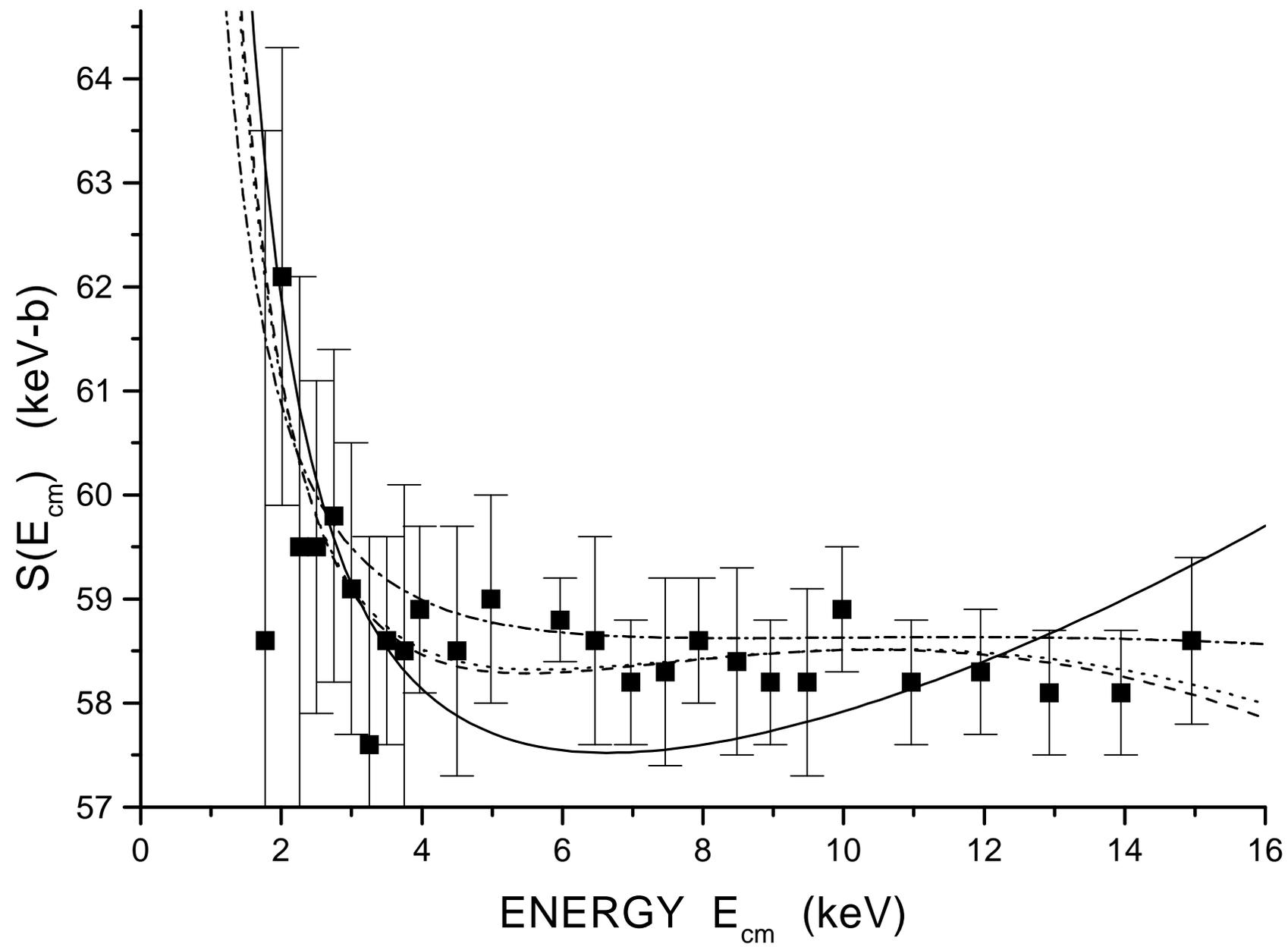
\end{document}